\documentclass[conference]{IEEEtran}

\usepackage{cite}
\usepackage{amsmath,amssymb,amsfonts}
\usepackage{graphicx}
\usepackage{multirow}

\def\BibTeX{{\rm B\kern-.05em{\sc i\kern-.025em b}\kern-.08em
    T\kern-.1667em\lower.7ex\hbox{E}\kern-.125emX}}
\newcommand{\cmmnt}[1]{\ignorespaces}

\makeatletter 
\newcommand{\linebreakand}{%
  \end{@IEEEauthorhalign}
  \hfill\mbox{}\par
  \mbox{}\hfill\begin{@IEEEauthorhalign}
}
\makeatother 

\begin{document}

\title{Keep what you need : extracting efficient subnetworks from large audio representation models
}

\author{\IEEEauthorblockN{David Genova}
\IEEEauthorblockA{
\textit{UMR 9912 STMS}\\
\textit{IRCAM - Sorbonne University - CNRS} \\
Paris, France \\
genova@ircam.fr}
\and
\IEEEauthorblockN{Philippe Esling}
\IEEEauthorblockA{
\textit{UMR 9912 STMS}\\
\textit{IRCAM - Sorbonne University - CNRS} \\
Paris, France \\
esling@ircam.fr}
\and
\IEEEauthorblockN{Tom Hurlin}
\IEEEauthorblockA{\textit{Squarp Instruments} \\
Paris, France \\
tom@squarp.net}
}

\maketitle

\begin{abstract}
Recently, research on audio foundation models has witnessed notable advances, as illustrated by the ever improving results on complex downstream tasks. Subsequently, those pretrained networks have quickly been used for various audio applications. These improvements have however resulted in a considerable increase both in size and complexity of these models. Along the environmental concerns this issue raises, this prevents the deployment of such networks on consumer-level devices, and precludes their use for real-time applications. Moreover, this appears contradictory with the specificity of the tasks for which these models are used, which are often simpler compared to extracting a rich, multi-purpose representation from any type of audio data. In this paper, we address this issue with a simple, yet effective method to extract lightweight specialist subnetworks from large foundation models. Specifically, we introduce learnable binary masks in-between the layers of a pretrained representation model. When training the end-to-end model on a downstream task, we add a sparsity-inducing loss to the overall objective, hence learning a compact subnetwork specialized on a single task. Importantly, the weights of the foundation model are kept frozen, resulting into low additional training costs. Once trained, the masked computational units can then be removed from the network, implying significant performance gains. We assess our method on three widespread audio foundation models, each based on a different backbone architecture, and illustrate its effectiveness on common audio representation evaluation tasks, as well as its versatility on both speech, music, and general audio. Code for reproducing the results and supporting webpage are available at https://github.com/gnvIRCAM/Audio-representation-trimming.
\end{abstract}
\section{Introduction}
\label{sec:intro}
Recent years have seen the emergence of deep neural networks as powerful and flexible solutions to address complex tasks in various domains. The development of \textit{foundation models} has allowed to extract rich representations from unlabelled data, which have found numerous applications in several fields, such as natural language processing \cite{devlin2019bert} or computer vision \cite{radford2021clip}. In audio, these models have led to significant progress in multiple areas of research, from audio information retrieval \cite{yuan2023marble} to generative modelling \cite{agostinelli2023musiclm}. After pre-training these models on vast corpuses of sounds taken from various audio domains, they are then adapted to downstream tasks on smaller datasets with limited data available, where fitting a randomly initialized network would likely result in overfitting. Hence, models such as HuBERT \cite{hsu2021hubert}, CLAP \cite{elizalde2023clap}, or MERT \cite{yizhi2023mert} have demonstrated state-of-the-art performance in transfer learning on speech, music, and general audio-related tasks, e.g. speech recognition, music auto-tagging, or environmental sounds classification. \\ 
However, foundation models have necessitated a continuous increase both in terms of number of parameters and computational complexity, which seems compulsory to encode sounds from highly variable audio domains into informative representations. Notably, this prevents their use for embedded applications on devices with limited computational resources. Moreover, despite being trained on simpler tasks with homogeneous data, the resulting models have even more parameters, due to the classification head, as well as potential additional modules \cite{adapter-network}. Hence, it is reasonable to assume smaller subnetworks could be extracted from the large foundation model, by keeping only the parts relevant for a single task, on a restrained set of sounds. \\
In order to mitigate such computational burdens, extensive research has been carried on neural network compression. 
Among existing methods, network \textit{pruning} \cite{lecun1989optimal} consists in identifying the less important weights of a network, which can then be removed without altering the performance of the model. While many pruning approaches allow for a substantial reduction in number of parameters \cite{frankle2018lottery, esling2020diet}, they usually require either fine-tuning or retraining the full model to achieve high compression ratios, hence being computationally expensive, and requires to have access to the original training dataset, which might not be publicly available. In this paper, we address the \textit{structured pruning} of audio representation models in a computationally-efficient way. Specifically, we introduce learnable binary masks after each intermediate layer of a pretrained foundation model. When training the model on a downstream task, we add a sparsity-inducing loss to the overall objective, hence learning a compact subnetwork specialized on a single task. Importantly, the weights of the foundation model are kept frozen, resulting into lower training costs than fine-tuning. Once trained, the masked computational units can then be removed from the network, thus implying inference speed-up and disk size reduction. We assess our method on three widespread audio foundation models and illustrate its effectiveness on common audio representation evaluation tasks for both music, speech, and general audio.

\section{Proposed approach}
Our goal is to remove computational blocks (i.e. columns for linear layers, convolution channels and attention heads) in large foundation models when training them on downstream tasks with smaller datasets. To this end, we introduce \textit{learnable masks} within the representation model, and introduce a \textit{sparsity-inducing loss} to the overall objective.  
\begin{figure*}
    \centering
    \includegraphics[scale=.8]{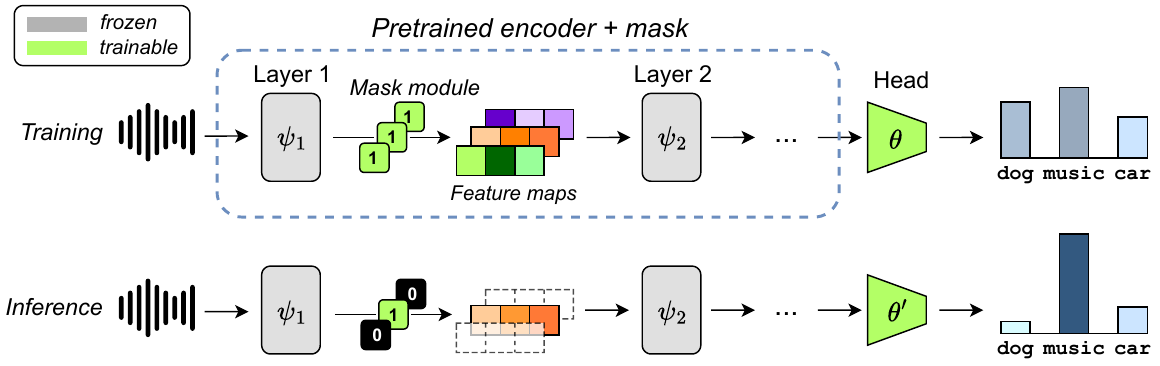}
    \caption{Overview of our approach. Given a frozen foundation model, each intermediate feature maps is processed by a learnable mask layer. After training, masked features correspond to unused computational units, which can be removed from the model.}
    \label{fig:approach}
\end{figure*}
\subsection{Foundation models}

\textit{Representation models} are a class of neural networks that aim at extracting compact embeddings from high-dimensional data. These models are usually trained in a \textit{self-supervised} way by either building discriminable representations using contrastive learning \cite{contrastiveaudio}, or by enforcing predictability of the embeddings in mask token modelling \cite{yizhi2023mert}. We base our work on three audio foundation models, each being trained on distinct audio domains (music, sound events or speech) and with different architectures (convolutional, transformer and conformer). Specifically, CLAP \cite{elizalde2023clap} is a multimodal network trained on paired text-audio data from multiple datasets \cite{fsd50k, clotho, audiocaps, macsdataset}. Using contrastive learning, the model builds a joint embedding space in which matching texts and sounds are encoded into aligned projections. Authors experiment with two different architectures, one using a 4-layer transformer network identical to HTSAT \cite{chen2022htsat}, the other being a 14-layer convolutional network introduced in PANNs \cite{kong2020panns}.
Whereas CLAP embeds an entire sound into a single vector, MusicFM \cite{musicfm} and Wav2Vec 2.0 \cite{baevski2020wav2vec} predict temporal sequences of codes. Both models are trained using masked token modeling, by randomly masking parts of the input, and then training the network to either predict the masked values or to identify them from a set of negative samples. MusicFM relies on the architecture of BEST-RQ \cite{bestrq} (12 conformer layers \cite{gulati2020conformer}), and is trained on musical data from both Free Music Archive \cite{musicfm} and an internal dataset. Wav2Vec 2.0 is composed of a 7 layers deep convolutional feature extractor, followed by 12 or 24 (base/large) transformer layers, and is trained on speech data from either Librispeech \cite{panayotov2015librispeech} or LibriVox \cite{kearns2014librivox}. As both Wav2Vec2 and MusicFM rely on a transformer backbone, we chose to use the convolutional CLAP model to cover a broader scope of architectures. 

\subsection{Transfer learning strategies}
Once trained to extract informative representations, foundation models can then be trained on more specific tasks, usually by adding a smaller network (such as a multi-layer perceptron or a recurrent network), denoted as the \textit{head}, at the end of the encoder. The resulting model is then trained using either \textit{linear probing}, which tunes only the parameters of the head, or \textit{fine-tuning}, where all the weights of the model are trained. Whereas linear probing is more compute-efficient, but can lead to subpar performances, fine-tuning often yields better results, but involves higher training costs and is more prone to overfitting. Hybrid approaches can be found in \textit{parameter-efficient transfer learning} (PETL), which provides both flexible and efficient strategies to transfer the knowledge of pretrained models. \textit{Adapter}-based methods \cite{adapter-network} introduce processing modules in-between the layers of foundation models. Closer to our work, \textit{Scaling-and-Shifting your Features} (SSF) \cite{lian2022scaling} adds learnable affine transforms modulating the intermediate features of the encoder, namely $\Tilde{f}_{(l, \theta)}(x) = \gamma_l \odot f_{(l, \theta)}(x) + \beta_l$, where $ \gamma_l, \beta_l \in \mathbb{R}^{d_\text{out}}$ are respectively \textit{scale} and \textit{shift} coefficients, 
$f_{(l, \theta)}$ is the \textit{l}-th layer of the encoder, and $d_\text{out}$ is the embedding dimension of $f_{(l, \theta)}$. Hence, our approach can be seen as a variant of SSF by enforcing $\beta_l=0$ and $\gamma_l \in \{0, 1\}^{d_\text{out}}$, while allowing reduction of the model by removing zeroed units once the training is complete. \\
Despite all of the benefits of representation learning, foundation models tend to be notably large, as they must have enough capacity to build rich representations for highly diverse data. Hence, such networks can not be used for offline applications, especially on embedded devices with limited computational resources.   

\subsection{Model compression}
Ever since the advent of large neural networks, researchers have dedicated efforts to make models smaller and faster. Early works on neural network \textit{pruning} \cite{lecun1989optimal, hassibi1993optimal} already showed that smaller networks can be found within larger ones, with improved generalization and reduced training costs. Follow-up research \cite{frankle2018lottery, frankle2020linear} established that most of the parameters of overparameterized networks could be removed (up to 95\%) while reaching equivalent (often even higher) accuracy. However, identifying these subnetworks requires multiple training cycles, and lead to sparse weight matrices which can not be converted into computational speed-up. Follow-up research proposed several approaches to prune networks in a single training run \cite{youdrawing, lee2019snip}, as well as removing parameters in a \textit{structured} fashion \cite{esling2020ultra}, thus truly leading to model reduction. However, these approaches do not match the performance of those relying on repeated training cycles, especially for structured pruning, which produces lower compression ratios than unstructured variants at matching accuracy.    

\subsection{Our approach}
Given a pretrained foundation model, we consider linear probing of the network (i.e. keeping its weights frozen) on a given downstream task. In this context, we hypothesize that many units of the model can be removed, while keeping a sufficiently informative representation for the considered task and dataset. Indeed, we believe that, while models with large capacity are necessary to build informative representations of audio data, downstream tasks do not leverage all of the available information contained in these embeddings, hence could be solved with a well-chosen subset of the network units.  \\ 
We define the end-to-end model as $f(x) = g(e(x, \psi), \boldsymbol{\theta})$,
where $g(., \boldsymbol{\theta})$ is the classification head and $e(., \psi)$ is the encoder (trainable parameters are indicated in bold). We focus on structural units of the encoder that can be entirely removed, such as convolution channels or attention heads, and note $l(., \psi_i), i \in \{1, ..., L\}$ the computational blocks composing the encoder, with $L$ being the number of such blocks. For these layers, we attach a trainable mask $\boldsymbol{m_i}$ such that $l(x, \psi_i, \boldsymbol{m_i})=\boldsymbol{m_i} \odot l(x, \psi_i) $. During the forward pass, the parameters of the mask are quantized into 0 or 1 using $l(x, \theta_i, \boldsymbol{m_i})=\textit{round}(\textit{sigmoid}(\boldsymbol{m_i}))l(x, \theta_i)$. Hence, each output feature of a computational unit can either be masked or kept. During back-propagation, we circumvent the non-differentiable rounding operation by using the straight-through estimator. By denoting $L_{C}$ the loss used for the downstream task, we add a sparsity-inducing loss to the overall objective $L_S = \sum_{i=1}^{N} \rVert \textit{sigmoid}(m_{i} - t)\rVert_2/N$, where $N$ is the total number of units and $t$ is an hyperparameter controlling the strength of the penality, with low values of $t$ pushing the network towards more sparsity. Together, the full objective can be written as $L=L_{C}+\lambda L_S$, where $\lambda$ is the weight of the sparsity loss. After training, the masked units can be removed from the model, leading to disk size reduction and computational speedup. 
\begin{table*}
    \caption{Results of our method across each models and datasets. For trimmed models, we indicate the evaluation metrics when removing resp. 25\%, 50\%, and 75\% of the parameters.}
    \centering
\begin{tabular}{c|c|c||c|c|ccc||ccccc}
\hline
\multirow{3}{*}{Model} & \multirow{3}{*}{Dataset} & \multirow{3}{*}{Metric} & \multirow{3}{*}{Base} & \multirow{3}{*}{SSF} & \multicolumn{3}{c||}{Scratch}  & \multicolumn{5}{c}{Our approach} \\ 
\cline{6-13} 
&  &  & \multicolumn{1}{|c|}{} & & \multicolumn{1}{c|}{\multirow{2}{*}{25\%}} & \multicolumn{1}{c|}{\multirow{2}{*}{50\%}} & \multirow{2}{*}{75\%} & \multicolumn{1}{c|}{\multirow{2}{*}{25\%}} & \multicolumn{1}{c|}{\multirow{2}{*}{50\%}} & \multicolumn{1}{c|}{\multirow{2}{*}{75\%}} & \multicolumn{2}{c}{Best} \\ \cline{12-13} 
&  &  &  & & \multicolumn{1}{c|}{} & \multicolumn{1}{c|}{} & & \multicolumn{1}{c|}{} & \multicolumn{1}{c|}{} & \multicolumn{1}{c|}{} & \multicolumn{1}{c|}{Trim. \%} & Score \\ 
\hline
\hline
\multirow{3}{*}{CLAP} & ESC50 & w-Acc. & .918  & \textbf{.951} & .733 & .711 & .719 & .910 & .911 & \multicolumn{1}{c|}{.851} & \multicolumn{1}{c|}{37.4} & \underline{.913} \\
     & US8K  & w-Acc. & .781  & \textbf{.860} & .661 & .650 & .632 & \underline{.838} & .823 & \multicolumn{1}{c|}{.791} & \multicolumn{1}{c|}{37.4} & .838 \\ 
     & FSD & mAP. & \underline{.483}  & \textbf{.489} & .291 & .279 & .289 & .478 & .448 & \multicolumn{1}{c|}{.239} & \multicolumn{1}{c|}{25.3} & .478 \\ 
\hline
\hline
\multirow{3}{*}{MusicFM}  & MTT-50 & mAP & \underline{.388} & .340 & .366 & .324 & .351 & \textbf{.393} & .380 & \multicolumn{1}{c|}{.362} & \multicolumn{1}{c|}{35.2} & \textbf{.393} \\
                          & GTZAN  & w-Acc & .868 & \textbf{.907} & .856 & .842 & .846 & .874 & .879 & \multicolumn{1}{c|}{.832} & \multicolumn{1}{c|}{40.8} & \underline{.888} \\ 
                          & NSynth  & w-Acc & .737 & .789 & .747 & \underline{.819} & \textbf{.830} & .787 & .785 & \multicolumn{1}{c|}{.731} & \multicolumn{1}{c|}{60.6} & .792\\ 
\hline
\hline
\multirow{3}{*}{Wav2Vec2} & FluentSpeech & w-Acc. & .802 & \textbf{.947} & .878 & .871 & .869 & \underline{.946} & .936 & \multicolumn{1}{c|}{.868} & \multicolumn{1}{c|}{25.0} & .946 \\
                          & LibriSpeech & WER (\%) & \underline{3.83} & \textbf{3.41} & 6.43 & 42.8 & 43.5 & 5.28 & 11.3 & \multicolumn{1}{c|}{33.9} & \multicolumn{1}{c|}{2.9} & 3.62 \\
                          & SpeechCommands & w-Acc. & .924 & \textbf{.970} & .878 & .935 & .941 & \underline{.960} & .958 & \multicolumn{1}{c|}{.946} & \multicolumn{1}{c|}{43.2} & .960
\end{tabular}
\label{table:results}
\end{table*}

\section{Experiments}
\label{sec:typestyle}
\paragraph*{Datasets and tasks}
We chose 9 publicly available datasets for downstream evaluation, corresponding to 3 tasks per audio domain (speech, sound event, music). For each dataset, we follow the official training/validation/test splits.\\
\textbf{\textit{Librispeech}} \cite{panayotov2015librispeech} is a corpus of audiobooks sampled at 16kHz, which we use for Automatic Speech Recognition (ASR).\\
\textbf{\textit{Fluent Speech Commands}} \cite{fluentspeech} is a dataset of 16kHz recordings designed for Intent Classification (IC). It is composed of 23,000 audios, each utterance corresponding to a specific action, object, and location.\\
\textbf{\textit{Speech Commands}} \cite{googlespeechcommands} comprises 65,000 1-second audio from thousands of speakers, used for keyword-spotting.\\
\textbf{\textit{ESC50}} \cite{esc50} is a dataset of 2000 environmental sounds, distributed in 50 classes (e.g. dog, vacuum cleaner or human laugh), each recording being 5 seconds long.\\
\textbf{\textit{Urbansound8k}} \cite{urbansound} is a collection of 8372 4 seconds-long samples from 10 classes (such as gunshot or jackhammer) for sound event classification.\\
\textbf{\textit{FSD50k}} \cite{fsd50k} is a sound event dataset for multilabel classification, in which each sound is associated to a set of labels from 200 diverse categories (notably including speech and music).\\
\textbf{\textit{GTZAN}} \cite{gtzan} is a dataset composed of 1000 30-seconds-long tracks, divided into 10 musical genres. To avoid biases from the original dataset, we use the fault-filtered split\footnote{https://github.com/coreyker/dnn-mgr/tree/master/gtzan}.\\
\textbf{\textit{MagnaTagTune}} \cite{magntagtune} is composed of more than 25,000 30-seconds-long musical clips labelled from a set of 188 tags. We keep the 50 most frequents ones for music auto-tagging. \\
\textbf{\textit{NSynth}} \cite{nsynth} is a dataset of 305,979 musical notes sampled at 16kHz, which we use for pitch estimation. \\
For evaluation metrics, we use weighted-accuracy (w-Acc.) for classification tasks, mean-Average precision (mAP) for auto-tagging, and word-error rate (WER) for ASR. 

\paragraph*{Implementation details}
We adapt the architecture of the head to the downstream task. Specifically, for classification and auto-tagging, we use a 2-layer MLP with hidden dimension 1024 and ReLU activation. For ASR, we add a 2-layer BiLSTM with hidden dimension 256. During training, the weights of the encoder are frozen and we train only the probing head and the masks. All models are trained for 100,000 steps using the Adam optimizer and a learning rate of 1e-3, using cross-entropy loss for classification, binary cross-entropy for music auto-tagging, and CTC loss for speech recognition. We set the weight of the sparsity loss to match the magnitude of the task objective. For each model, we report the results for trimming ratios (i.e number of parameters removed from the network) of $25\%$, $50\%$ and $75\%$, which corresponds to different values of $t \in [0.3, 0.7]$ for each model and task. 

\paragraph*{Baselines and ablations}
For each model, we train a model of equivalent size from scratch, with the same capacity as the trimmed model. Specifically, given a specific task, we use the mask obtained using our method to trim the foundation model. We then re-initialize the remaining weights and train the newly obtained model from scratch. For MusicFM and Wav2Vec2, we apply learning rate warmup for 25,000 steps, which is known to play a critical role when training transformers, and lower the final learning rate value to 1e-4. We also evaluate our approach against the SSF approach, by removing the sparsity loss and replacing the mask modules with modulation layers. 

\section{Results}
\label{sec:results}

\subsection{Impact of sparsity ratio on model performance}

In Table \ref{table:results}, we present the results of our approach for the aforementioned models and datasets. We found that, for most models, our approach \textit{outperforms} regular linear probing up to 25\% of removed parameters, and still nearly matches the performance of the full model for even larger sparsity ratios (40-50\%). For most tasks, our method consistently produces better results than training a network of equivalent size from scratch (which almost always results in overfitting), while being significantly cheaper to train. In most cases, SSF outperforms our approach on accuracy, which is to be expected as this approach can adapt the network features with more flexibility than ours. Yet, SSF does not lead to any reduction of the model. When training Wav2Vec2 for ASR, we note performance drops at relatively small trimming ratios (compared to the other models and tasks), which can be attributed to the complexity of this task (being both a classification and sequence alignment problem). Yet, the trimmed networks still reach competitive performance (7.2\% WER for 40\% sparsity) and could still be used for efficient ASR with only a small accuracy loss. We note a similar phenomenon with CLAP for mutilabel classification on FSD, where trimming the model always results in lower mAP, which can be explained by the high diversity of sounds and labels within the dataset. Generally, we note that the optimal trimming ratio (i.e. number of removed units associated to the best evaluation score) is strongly dependent on the task. Indeed, as illustrated in the two rightmost columns of Table \ref{table:results}, models trained for complex tasks (e.g. seq2seq, multiple labels or diverse dataset) will require larger subnetworks than when trained on simpler tasks such as single label classification.            

\subsection{Computational gain}
In Table \ref{table:computation}, we present the computational gains when trimming foundation models using our method. As these gains will depend on the percentage of removed units, for each model, we set a maximal performance drop of 5\% (relatively to the full model), and report the metrics for the fastest model within this range and averaged across \textit{tasks}, namely classification (classif.), audio tagging and ASR. Hence, the values in Table \ref{table:computation} should be read as \textit{expected} gains for tasks of comparable complexity. 

\begin{table}[h]
    \caption{Expected computational gains for each model. base and trim. stand for the original and trimmed model. } 
    \centering
    \begin{tabular}{c|c|c|c|c|c} 
    \hline
    \multicolumn{2}{c|}{Model} &Size (Mo)& GFLOPs & GMACs &Speedup        \\ 
    \hline\hline
    \multirow{3}{*}{CLAP}     &base& 344.2     & 23.3  & 11.6 & -     \\ 
                              &trim. (classif.)& 95.9      & 7.6   & 3.8  & $\times 2.8$   \\ 
                              &trim. (tagging)& 198.1      & 14.8   & 7.4  & $\times 1.4$   \\ 
    \hline\hline
    \multirow{3}{*}{MusicFM}  &base& 836.6     & 144.4 & 72.2 & -    \\ 
                              &trim. (classif.)& 249.9  & 114.8 & 57.4 & $\times 1.4$  \\ 
                              &trim. (tagging)& 230.8     & 113.9 & 56.9 & $\times 1.3$  \\ 
    \hline\hline
    \multirow{3}{*}{Wav2Vec2} &base& 380.9     & 55.5  & 27.7 & -    \\ 
                              &trim. (classif.)& 89.3     & 13.4  & 6.7  & $\times 2.5$   \\
                              &trim. (ASR)& 233.5     & 36.8  & 17.9  & $\times 1.2$   \\
    \end{tabular}
    \label{table:computation}
\end{table}

For CLAP and Wav2Vec 2.0, trimming significantly reduces the disk size and computational complexity of the model, resulting in faster inference. For MusicFM, even though the size of the model is also reduced, the influence of trimming on the number of operations, hence on speedup, is less significant. We attribute this gap to either an intrinsic bottleneck of the implementation we used \footnote{https://github.com/minzwon/musicfm}, or to the architecture of the model.

\section{Conclusion}
\label{sec:conclusion}
In this paper, we showed that foundation models can be drastically reduced when used in downstream tasks setups. We illustrated our approach in a wide variety of cases and different architectures and found that, for most cases, a smaller model can achieve similar (and even in some cases \textit{better}) performance than the full network. We also found that trimmed models effectively yield faster inference and decreased disk size. Hence, our approach provides a step towards making such models suitable for on-device applications. For future works, we seek to extend our method to a wider set of applications, notably for audio generative modelling, which has greatly benefited from the improvements in representation learning. Moreover, we seek to better understand the structure of these learned masks, and how we could leverage them to better understand the features learned by audio foundation models. 

\newpage
\bibliographystyle{IEEEbib}
\bibliography{strings,refs}

\end{document}